\documentclass[showpacs,amsmath,amssymb,aps,floatfix,nofootinbib,twocolumn]{revtex4}
\usepackage[dvips]{graphicx}% Include figure files
\usepackage{dcolumn}% Align table columns on decimal point
\usepackage{bm}% bold math
\usepackage[caption=false]{subfig}
\usepackage{slashed,epsfig}
\usepackage{amsthm,amssymb,amsfonts,multirow,array}
\usepackage[colorlinks=true,citecolor=blue,urlcolor=blue,linkcolor=blue]{hyperref}
\usepackage{mathtools}
\usepackage{xcolor,appendix}
\usepackage{float, physics}
\usepackage[utf8]{inputenc}
\usepackage{amsmath}
\begin{document}
\preprint{APS/123-QED}
\title{Spectroscopic Analysis of Hidden-Charm Pentaquarks}
\author{\large Ankush Sharma}
\thanks{ankushsharma2540.as@gmail.com}
\author{\large Alka Upadhyay}

\affiliation{\small School of Physics and Materials Science, Thapar Institute of Engineering and Technology, Patiala - 147004, Punjab, INDIA}
\date{\today}% It is always \today, today,
\baselineskip =1\baselineskip

\begin{abstract}
In this work, the multiquark approach is used to analyze the spectroscopy of hidden-charm pentaquark states, motivated by recent discoveries at the LHCb collaboration. Using the SU(3) flavor representation, pentaquarks having $J^P = 5/2^-$  are arranged into 10 (decuplet) of the SU(3) flavor multiplets. The masses of pentaquarks are calculated using the extension of the Gursey-Radicati mass formula and the effective mass scheme. Also, we calculated the magnetic moments of the hidden-charm pentaquarks using the effective mass and shielded charge technique. Further, we suggested the possible production modes for $J^P = 5/2^-$ pentaquarks from the decay of bottom baryons, which consist of pentaquark states as intermediate states. Our results for masses demonstrate reasonable agreement with the available data and our analysis for both masses and magnetic moments may be useful for future experimental studies.
\end{abstract}

%\keywords{Multiquark particles, 14.20.-c Baryons; 14.65.Dw Charmed quarks; 12.39.-x Phenomenological quark models; 02.20.-a Group theory}%Use showkeys class option if keyword
\maketitle

%\tableofcontents

\section{Introduction}
%\label{sec:intro}
Recent studies from different experimental collaborations confirm the existence of exotic hadron states like tetraquarks, pentaquarks, hexaquarks, etc. Dozens of exotic states have been discovered in the past decade. In this work, we limit ourselves to study the hidden-charm pentaquark states by taking the motivation from the recent discoveries. Recently, in 2022, LHCb discovered two tetraquarks and hidden-charm pentaquark structures \cite{4338}. Pentaquark $P_{\psi s}^\Lambda(4338)^0$ having minimal quark content $udsc\overline{c}$, is observed in the $J/\psi\Lambda$ mass in the $B^-\rightarrow J/\psi\Lambda p$ decays with the statistical significance of 15$\sigma$, which is far beyond the five standard deviations that are required to claim the observation of a particle. The mass and the width of the new pentaquark are measured to be $4338.2\pm0.7\pm0.4$ MeV and $7.0\pm1.2\pm1.3$ MeV, respectively. The preferred quantum numbers are $J^P = 1/2^-$. Also, in 2021, the pentaquark state $P_{cs}(4459)$ was observed in the J/$\psi$ $\Lambda$ invariant mass distribution from an amplitude analysis of the $\Xi_b^- \rightarrow J/\psi \Lambda K^-$ decays \cite{20211278}. The observed structure is consistent with a hidden-charm pentaquark with strangeness, characterized by a mass of $4458.8\pm2.9_{-1.1}^{+4.7}$ MeV and a width of $17.3\pm6.5_{-5.7}^{+8.0}$ MeV. The spin is expected to be 1/2 or 3/2, and its parity can be either of $\pm 1$.
On the other hand, in 2019, LHCb reported three hidden-charm pentaquark structures, $P_c(4312)^+$ decaying to $J/\psi p$, is discovered with a statistical significance of 7.3$\sigma$ in a data sample of $\Lambda_b^0 \rightarrow J / \psi pK^-$ decays, $P_c(4450)^+$ pentaquark structure consists of two narrow overlapping peaks, $P_c(4440)^+$ and $P_c(4457)^+$, where the statistical significance of this two-peak interpretation is 5.4$\sigma$ \cite{PhysRevLett.122.222001}. Also, in 2015, the LHCb collaboration reported two hidden-charm structures, $P_c(4380)^+$, and $P_c(4450)^+$, in $\Lambda_b$ decay. These two pentaquark states have the mass $4380\pm8\pm28$ MeV and $4449.8\pm1.7\pm2.5$ MeV with corresponding widths of $205\pm18\pm86$ MeV and $39\pm5\pm19$ MeV respectively.  According to the LHCb collaboration, the preferred $J^P$ assignments for these states are $3/2^-$ and $5/2^+$, respectively \cite{PhysRevLett.115.072001}. These discoveries are further supported by another two articles of the same collaboration \cite{PhysRevLett.117.082002, PhysRevLett.117.082003}. These observations motivate theoreticians to study hidden-charm pentaquark states and associated properties. Many approaches have been proposed to explain hidden-charm pentaquarks, such as two colored diquarks bound with an antiquark \cite{MAIANI2015289,anisovich2015pentaquarks, wang2016analysis, wang2016analysis1}, and compact diquark model \cite{Rehman19, Li15}, diquark-triquark model \cite{LEBED2015454}, which describes cluster structure for quarks inside pentaquarks. 
 Also, hidden-charm pentaquarks can be interpreted as a molecular state having a charmed meson-charmed baryon molecular state, for example, $P_c(4380)^+$ and $P_c(4450)^+$ being $\overline{D}\Sigma_c^*$ and  $\overline{D}^*\Sigma_c$ molecular states respectively \cite{PhysRevLett.115.172001, CHEN2016406, PhysRevD.92.094003, HE2016547, burns2015phenomenology, PhysRevLett.115.122001, PhysRevLett.115.132002, MEINER201559, Bo19, PTEP}, as a soliton state pentaquarks \cite{PhysRevD.92.051501}, and, threshold enhancement or kinematic effect  \cite{PhysRevD.92.071502,LIU2016231,Mikhasenko:2015vca,Anisovich_2015}. Considering the recent discoveries of pentaquarks resonances by LHCb in $\Lambda_b^0 \rightarrow J/\psi pK^-$ as genuine states, which provides a rich source for the pentaquark production in bottom baryon decays since LHCb can produce a large number of bottom baryons.\\
Now, for the case of light quarks ($u$, $d$, $s$), we use the SU(3) flavor representation to classify the pentaquark states into flavor multiplets, with each quark assigned as fundamental 3 and an anti-quark by $\overline{3}$ representation. As a result, pentaquarks with the configuration  $qqqq\overline{q}$ are classified as: 
\begin{equation}
    [3] \otimes [3] \otimes [3] \otimes [3] \otimes
[\overline{3}] = 3[1] \oplus 8[8] \oplus 4[10] \oplus 2[\overline{10}] \oplus 3[27] \oplus [35]
\end{equation}
Thus, the pentaquark state can be in any of the flavor multiplets of SU(3) flavor representation. Further, ground state pentaquarks have negative parity ($l=0$) as the parity for the case of hidden-charm pentaquark system is defined as:
\begin{equation}
    P\vert{qqqc\overline{c}}\rangle = (-1)^{l+1}\vert{qqqc\overline{c}}\rangle
\end{equation}
Also, $c\overline{c}$ having spin 0 or 1 can be in a color singlet or octet; therefore remaining three light quarks must be in a color singlet or octet state respectively to make the total color wavefunction of the pentaquark color singlet. Further, to define the spin wave functions for pentaquarks, we are using the SU(2) spin representation where a fundamental 2 is assigned to each quark as well as to an anti-quark is:
\begin{equation}
    [2] \otimes [2] \otimes [2] \otimes [2] \otimes
[2] =  [6]  \oplus 4[4] \oplus 5[2]
\end{equation}
where the configurations [6], [4], and [2] correspond to the spin values $s$ = 5/2, 3/2, and 1/2, respectively. In this work, we are interested in the hidden-charm decuplet of pentaquark states having $J^P = 5/2^-$. Now, for the case of 5 flavors ($u$, $d$, $s$, $c$, $b$), hypercharge is defined as \cite{Workman:2022ynf}:
\begin{equation}
    Y = B + S - \frac{C - B' + T'}{3} 
\end{equation}
where $B$ is the baryon number, $S$ is the strangeness, $C$ is the charm, $B'$ is the bottom and $T'$ are the top quantum numbers for quarks and antiquarks. Since, there is a $c\bar{c}$ pair in hidden-charm pentaquarks, therefore, hypercharge $Y$ for three quarks system ($u$, $d$, $s$) is:
\begin{equation}
    Y = B + S
\end{equation}
Further, allowed $SU_f(3)$ multiplets to which charmonium pentaquark belongs are:\
 \begin{equation}
     [111]_1,\hspace{0.5cm} [21]_8,\hspace{0.5cm}   [3]_{10}
 \end{equation}
  We excluded the singlet $[111]_1$ as it contains no sub-multiplets with hypercharge $Y = 1$ \cite{PhysRevD.96.014014}. Therefore octet (8) and decuplet (10) are the remaining possible sub-multiplets of $SU_f(3)$ for charmonium pentaquarks. Assuming that the charmonium decuplet of pentaquarks may be found in $\Lambda_b^0$ decays, we have classified hidden-charm pentaquarks into $SU(3)_f$ decuplet ($J^P =5/2^-$) and calculated their masses using the extension of the Gursey-Radicati (GR) mass formula and the effective mass scheme.  GR mass formula is a useful tool for studying the mass spectrum of hadrons. In the past, many works have been proposed using the GR mass formula. In Ref.\cite{PhysRevD.96.014014}, the mass spectrum of the hidden-charm octet of pentaquarks has been calculated using the same formalism where they confirm one of the newly discovered pentaquark structure by LHCb .i.e. $P_c(4380)$ as a member of $SU(3)_f$ octet and in Ref. \cite{HOLMA2020135108}, different fits for mass formula parameters have been employed to calculate the masses of hidden-charm pentaquark states. Recently, in 2023, we have successfully computed the masses of tetraquark states using the extension of the GR mass formula \cite{Sharma_2023}. Therefore, the extension of the GR mass formula is a very useful method for the prediction of hadron masses. Also, magnetic moments of the predicted decuplet states are analyzed using the effective mass and screened charge technique, which will help us explore the inner structure of pentaquark states. Furthermore, we proposed the bottom baryon decay modes for the production of hidden-charm pentaquark states, which may be useful for future experimental studies. \\
This work is organized as follows: Section II briefly introduces the theoretical formalism which consists of the extension of the Gursey-Radicati mass formula and its application to find the masses of hidden-charm pentaquark states, effective mass, and shielded charge technique for masses and magnetic moments of pentaquarks. Section III discusses the production of pentaquark states via bottom baryon decay channels, which involves them as intermediate assignments. Section IV sums up the results and the conclusion.
\section{Theoretical Formalism}
\subsection{The Extended Gursey-Radicati Mass Formula}
It was F. Gursey, and L. Radicati, who invented the mass formula to study the mass spectrum of baryons \cite{PhysRevLett.13.173}:
\begin{align}
  M = M_0 + a J(J+1) + bY + c [T(T+1) - 1/4 Y^2]
\end{align}
This is the most general mass formula based on broken SU(6) symmetry. $J$ and $T$ are the spin and isospin quantum numbers, and $a$, $b$, and $c$ are the mass formula parameters. In 2004, the mass formula was generalized in terms of the Casimir operator of SU(3) representation as \cite{10.1063/1.1805936}:
\begin{align}
 M_B = M_0 + A S(S+1) + DY+ G C_2(SU(3)) \nonumber \\  + E [I(I+1) - 1/4 Y^2]
  \end{align}
  To study the mass splitting between the different multiplets of hidden-charm pentaquarks, we considered the extension of the Gursey-Radicati mass formula which distinguishes the different multiplets of SU(3) is \cite{PhysRevD.96.014014}: 
 \begin{align}
  M_{GR} = M_0 + AS(S+1) + DY  + E[I(I+1) -1/4 Y^2] \nonumber \\  + G C_2(SU(3)) + F N_c
  \end{align}

  where $M_0$ is a scale parameter related to the number of quarks in the system. $S$, $I$, and $Y$, are the spin, isospin, and hypercharge quantum numbers, respectively. $C_2(SU(3))$ stands for the eigenvalue of the $SU_f(3)$ Casimir operator and its value for different charmonium multiplets is written in Table \ref{tab:3}. $N_c$ stands for the counter of $c$ quarks or $\overline{c}$ anti-quarks. The parameters $A$, $D$, $G$, $E$, and $F$ are the mass formula parameters listed in Table \ref{tab:4} with their respective uncertainties. To calculate the mass formula parameters, well-known baryon spectra have been used as complete information about the pentaquark spectrum is still not available. We assumed that the value of these mass formula parameters ($A$, $D$, $G$, $E$, and $F$) are the same for all hadronic systems. The mass spectrum for the ground-state charmed baryons, hyperons, and non-strange baryons is listed in Table \ref{tab:1}. In contrast, Table \ref{tab:2} contains information about their associated quantum number assignments. These parameters are evaluated by fitting them together to obtain the best reproduction of the spectrum of ground-state baryons, charmed baryons, and non-strange baryons written in Table \ref{tab:4}.

\begin{table}[ht]
\centering
\begin{tabular}{ccc}
 \hline
 \hline
      Baryons & \hspace{0.3cm} Expt.
    Mass [MeV] & \hspace{0.3cm} Expt. error.\\
      \hline
        N(940) & \hspace{0.3cm} 939.565 & \hspace{0.3cm} $10^{-6}$ \\
        $\Lambda^0$(1116) & \hspace{0.3cm} 1115.68 & \hspace{0.3cm} 0.006 \\
$\Sigma^0$(1193) & \hspace{0.3cm} 1192.64 & \hspace{0.3cm} 0.024 \\
$\Xi^0(1315)$ & \hspace{0.3cm} 1314.86 & \hspace{0.3cm} 0.20 \\
$\Delta^0$(1232) & \hspace{0.3cm} 1232.00 & \hspace{0.3cm} 2.00 \\
$\Sigma^0$(1385) & \hspace{0.3cm} 1383.70 & \hspace{0.3cm} 1.00 \\
$\Xi^0$(1530) & \hspace{0.3cm} 1531.80 & \hspace{0.3cm} 0.32 \\
$\Omega^-$(1672) & \hspace{0.3cm} 1672.45 & \hspace{0.3cm} 0.29 \\
$\Lambda_c^+$(2286) & \hspace{0.3cm} 2286.46 & \hspace{0.3cm} 0.14 \\
$\Sigma_c^0$(2455) & \hspace{0.3cm} 2453.75 & \hspace{0.3cm} 0.14 \\
$\Xi_c^0$(2471) & \hspace{0.3cm} 2470.85 & \hspace{0.3cm} $^{+0.28}_{-0.40}$\\
$\Xi_c^{'0}$(2576) & \hspace{0.3cm} 2577.90 & \hspace{0.3cm} 2.90 \\
$\Omega_c^0$(2695) & \hspace{0.3cm} 2695.20 & \hspace{0.3cm} 1.70 \\
$\Omega_c^{*0}$(2770) & \hspace{0.3cm} 2765.90 & \hspace{0.3cm}  2.00\\
$\Sigma_c^{*0}$(2520) & \hspace{0.3cm} 2518.48 & \hspace{0.3cm} 0.20 \\
$\Xi_c^{*0}$(2645) & \hspace{0.3cm} 2649.90 & \hspace{0.3cm} 0.50 \\
\hline
\hline
    \end{tabular}
\caption{Mass spectrum of all the ground state charmed baryons, hyperons, and the non-strange baryons as from Particle data group \cite{Patrignani}.\label{tab:1}}
\end{table}

\begin{table}[!ht]
    \centering
 {\renewcommand{\arraystretch}{1.2}
    \begin{tabular}{ccccccc}
    \hline
    \hline
       Baryons & \hspace{0.5cm} $SU_f(3)$ & \hspace{0.5cm} $C_2(SU(3)$ & \hspace{0.5cm} S & Y & I & $N_c$ \\
       \hline
       N(940) & \hspace{0.5cm} $[21]_8$ & \hspace{0.5cm}3 & \hspace{0.5cm} $\frac{1}{2}$ & 1 & $\frac{1}{2}$ & 0 \\
$\Lambda^0(1116)$ & \hspace{0.5cm} $[21]_8$ & \hspace{0.5cm} 3 & \hspace{0.5cm} $\frac{1}{2}$ & 0 & 0 & 0 \\ 
$\Sigma^0(1193)$ & \hspace{0.5cm} $[21]_8$ & \hspace{0.5cm} 3 & \hspace{0.5cm} $\frac{1}{2}$ & 0 & 1 & 0 \\
$\Xi^0(1315)$ & \hspace{0.5cm} $[21]_8$ & \hspace{0.5cm} 3 & \hspace{0.5cm} $\frac{1}{2}$ & -1 & $\frac{1}{2}$ & 0 \\
$\Delta^0(1232)$ & \hspace{0.5cm} $[3]_{10}$ & \hspace{0.5cm} 6 & \hspace{0.5cm} $\frac{3}{2}$ & 1 & $\frac{3}{2}$ & 0 \\
$\Sigma^0(1385)$ & \hspace{0.5cm} $[3]_{10}$ & \hspace{0.5cm} 6 & \hspace{0.5cm} $\frac{3}{2}$ & 0 & 1 & 0 \\
$\Xi^0(1530)$ & \hspace{0.5cm} $[3]_{10}$ & \hspace{0.5cm} 6 & \hspace{0.5cm} $\frac{3}{2}$ & -1 & $\frac{3}{2}$ & 0 \\
$\Omega^{-}(1672)$ & \hspace{0.5cm} $[3]_{10}$ & \hspace{0.5cm} 6 & \hspace{0.5cm} $\frac{3}{2}$ & -2 & 0 & 0 \\ 
$\Lambda_c^+$(2286) & \hspace{0.5cm} $[11]_3$ & \hspace{0.5cm} $\frac{4}{3}$ & \hspace{0.5cm} $\frac{1}{2}$ & $+\frac{2}{3}$ & 0 & 1 \\ 
$\Sigma_c^0$(2455) & \hspace{0.5cm} $[2]_6$ & \hspace{0.5cm} $\frac{10}{3}$ & \hspace{0.5cm} $\frac{1}{2}$ & $+\frac{2}{3}$ & 1 & 1 \\
$\Xi_c^0$(2471) & \hspace{0.5cm} $[11]_3$ & \hspace{0.5cm} $\frac{4}{3}$ & \hspace{0.5cm} $\frac{1}{2}$ & $-\frac{1}{3}$ & $\frac{1}{2}$ & 1 \\ 
$\Xi_c^{'0}$(2576) & \hspace{0.5cm} $[2]_6$ & \hspace{0.5cm} $\frac{10}{3}$ & \hspace{0.5cm} $\frac{1}{2}$ & $-\frac{1}{3}$ & $\frac{1}{2}$ & 1 \\ 
$\Omega_c^0$(2695) & \hspace{0.5cm} $[2]_6$ & \hspace{0.5cm} $\frac{10}{3}$ & \hspace{0.5cm} $\frac{1}{2}$ & $-\frac{4}{3}$ & 0 & 1 \\
$\Omega_c^{*0}$(2770) & \hspace{0.5cm} $[2]_6$ & \hspace{0.5cm} $\frac{10}{3}$ & \hspace{0.5cm} $\frac{3}{2}$ & $-\frac{4}{3}$ & 0 & 1 \\
$\Sigma_0^{*0}$(2520) & \hspace{0.5cm} $[2]_6$ & \hspace{0.5cm} $\frac{10}{3}$ & \hspace{0.5cm} $\frac{3}{2}$ & $+\frac{2}{3}$ & 1 & 1 \\
$\Xi_c^{*0}$(2645) & \hspace{0.5cm} $[2]_6$ & \hspace{0.5cm} $\frac{10}{3}$ & \hspace{0.5cm} $\frac{3}{2}$ & $-\frac{1}{3}$ & $\frac{1}{2}$ & 1 \\
\hline
\hline
    \end{tabular}
    }
     \caption{Quantum number assignments for the ground state baryons mentioned in Table \ref{tab:1}. $S$, $I$, $Y$ are the spin, isospin and hypercharge quantum numbers respectively. $N_c$ stands for the number of charm quarks.}
     \label{tab:2}
   \end{table} 
   
\begin{figure}
    \centering
    \includegraphics[width=0.8\linewidth]{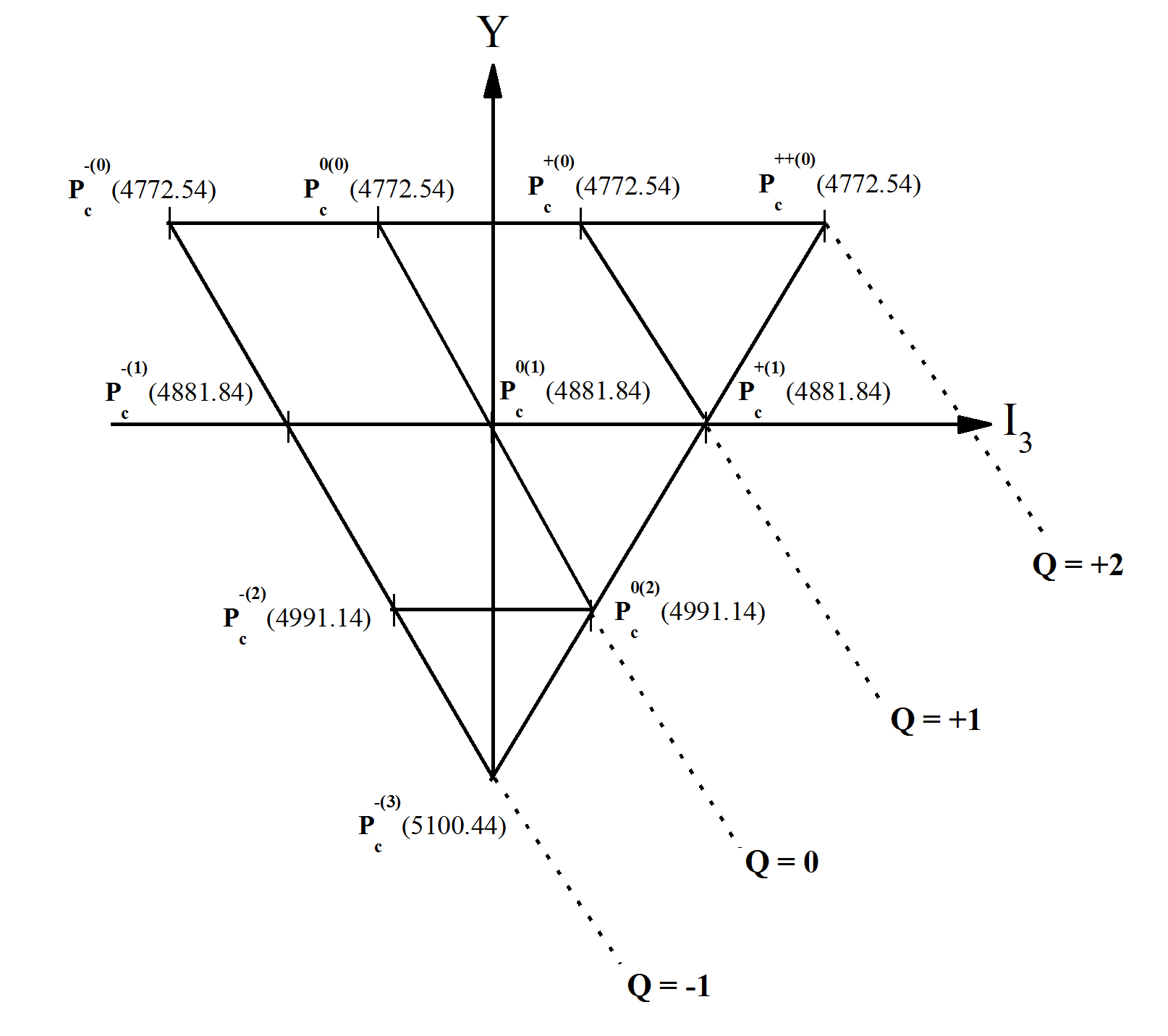}
    \caption{Hidden-charm decuplet of pentaquarks having $J^P = 5/2^-$, Each state is labelled as $P_c^{q(s)}$, where $q$ is the electric charge and $s$ is the strangeness of the state.}
    \label{fig:1}
\end{figure}

 \begin{table}[ht]
       \centering
       \begin{tabular}{cc}
       \hline
       \hline
         $SU_f(3)$ multiplet & \hspace{0.5cm} $C_2(SU(3))$ \\
           \hline
        $[3]_{10}$ & \hspace{0.5cm} 6 \\
         $[21]_{8}$ & \hspace{0.5cm} 3 \\
        \hline
        \hline
       \end{tabular}
        \caption{Possible charmonium pentaquark multiplets with their corresponding value of Casimir operator $C_2(SU(3))$ \cite{PhysRevD.96.014014}.}
        \label{tab:3}
   \end{table}
 We used the extended GR mass formula to one of the possible $SU_f(3)$ multiplet of the charmonium states, i.e., hidden-charm decuplet of pentaquarks. It is applied to each state of the decuplet to predict their corresponding masses. The decuplet structure of pentaquarks is reported in Figure \ref{fig:1}, while the predicted masses with the related uncertainties are listed in Table \ref{tab:5}, and a comparison of the mass spectra is carried out with the available theoretical data. Each pentaquark state is denoted as $P_c^{q(s)}(M)$, where $q$ is the electric charge, $s$ stands for the strangeness and $M$ is the predicted mass of the particles. 

\begin{table}[ht]
       \centering
       \begin{tabular}{ccccccc}
       \hline
       \hline
         & $M_0$ & A & D & E & F & G \\
         \hline
        Values[MeV] & 940.0 & 23.0 & -158.3 & 32.0 & 1354.6 & 52.5 \\
        Uncertainties [MeV] & 1.5 & 1.2 & 1.3 & 1.3 & 18.2 & 1.3 \\
        \hline
        \hline
       \end{tabular}
        \caption{Values of parameters used in the extended GR mass formula with corresponding uncertainties \cite{PhysRevD.96.014014}.}
        \label{tab:4}
   \end{table}

  \begin{table*}[ht!]
        \centering
          \begin{tabular}{cccc}
      \hline
      \hline
   Pentaquark States & \hspace{0.3cm}  Masses [MeV] & \hspace{0.3cm} Effective Mass Scheme [MeV]  & \hspace{0.3cm} Ref. \cite{isomers} [MeV]\\
     \hline
     $P_c^{++(0)}$, $P_c^{+(0)}$,  $P_c^{0(0)}$, $P_c^{-(0)}$ & \hspace{0.3cm} 4772.54 $\pm$ 38.99   & \hspace{0.3cm} 4615.64 & \hspace{0.3cm} 4743\\ 
     $P_c^{+(1)}$, $P_c^{0(1)}$, $P_c^{-(1)}$  & \hspace{0.3cm} 4881.84 $\pm$ 38.79  & \hspace{0.3cm} 4772.42 & \hspace{0.3cm} 4850 \\ 
     $P_c^{0(2)}$, $P_c^{-(2)}$ &  \hspace{0.3cm} 4991.14 $\pm$ 38.73   & \hspace{0.3cm} 4931.88 & \hspace{0.3cm} 5008\\ 
     $P_c^{-(3)}$ & \hspace{0.3cm} 5100.44 $\pm$ 38.81  & \hspace{0.3cm} 5094.02 & \hspace{0.3cm} 5140 \\ 
     \hline
     \hline
       \end{tabular}
        \caption{Table for masses of predicted Pentaquark Decuplet. The notation for pentaquark states are same as that of Fig.\ref{fig:1}.}
    \label{tab:5}
\end{table*}

\subsection{Effective Mass Scheme}
The magnetic moment encoding of the pentaquark provides valuable information regarding the distribution of charge and magnetization within the hadrons, thereby aiding in the analysis of their geometric configurations. In this work, we computed the effective mass of quarks (antiquarks) by considering their interaction with neighboring quarks through one gluon exchange scheme. By using effective quark masses, we calculated the masses and magnetic moments of hidden-charm pentaquarks, which helps us to explore their inner structure. The mass of pentaquarks can be written as \cite{VERMA} :
\begin{equation}
    M_P = \sum_{i=1}^5 m_i^\epsilon = \sum_{i=1}^5 m_i + \sum_{i<j} b_{ij} s_i.s_j
    \label{10}
\end{equation}
 here, $s_i$ and $s_j$ represent the spin operator for the $i^{th}$ and $j^{th}$ quarks (antiquark) and $m_i^\epsilon$ represents the effective mass for each of the quark (antiquark) and $b_{ij}$ is defined as:
 \begin{equation}
     b_{ij} = \frac{16 \pi \alpha_s}{9 m_i m_j} \bra{\Psi_0}\delta^3(\Vec{r})\ket{\Psi_0}
 \end{equation}
 where $\Psi_0$ is the ground state pentaquark wavefunction.
 For different quarks inside the pentaquark, effective masses equations are:
 \begin{equation}
 m_1^{eff} = m_1 + \alpha b_{12} + \beta b_{13} + \gamma b_{14} + \eta b_{15}
 \end{equation}
 
 \begin{equation}
  m_2^{eff} = m_2 + \alpha b_{12} + \beta^{'} b_{23} + \gamma^{'} b_{24} + \eta^{'} b_{25}
  \end{equation}

   \begin{equation}
   m_3^{eff} = m_3 + \beta b_{13} + \beta^{'} b_{23} + \gamma^{''} b_{34} + \eta^{''} b_{35} 
   \end{equation}

   \begin{equation}
    m_4^{eff} = m_4 + \gamma b_{14} + \gamma^{'} b_{24} + \gamma^{''} b_{34} + \eta^{'''} b_{45} 
 \end{equation}

  \begin{equation}
    m_5^{eff} = m_5 + \eta b_{15} + \eta^{'} b_{24} + \eta^{''} b_{34} + \eta^{'''} b_{45} 
 \end{equation}
  Here, 1, 2, 3, 4, and 5 stand for $u$, $d$, $s$, $c$, and $b$ quarks. These equations get modified if we consider two/three/four/five identical quarks. Therefore, products of spin quantum number are defined as \cite{Rohit}:
\begin{equation}
s_1.s_2 = s_2.s_3 =  s_3.s_4 =
 s_4.s_5 = 1/4 
\end{equation}
 and parameters are calculated by using eq.\eqref{10}:
 \begin{equation}
     \alpha = \beta = \gamma = \eta = 1/8
 \end{equation}

\begin{equation}
\beta^{'} = \gamma^{'} = \eta^{'} = 1/8
\end{equation}

\begin{equation}
\gamma^{''} = \eta^{''} = \eta^{'''} = 1/8
\end{equation}
 The value of the quark masses are taken from Ref. \cite{Rohit} and hyperfine interaction terms $b_{ij}$ are calculated using the effective mass equations and written as:

\begin{align}
     m_u = m_d = \hspace{0.3cm} 362 MeV, \hspace{0.3cm} m_s = \hspace{0.3cm} 539 MeV \nonumber \\
    m_c = \hspace{0.3cm} 1710 MeV, \hspace{0.3cm} m_b = \hspace{0.3cm} 5043 MeV
\end{align}

 \begin{equation}
     b_{uu} = \hspace{0.3cm} b_{ud} = \hspace{0.3cm} b_{dd} = 101.66 MeV
 \end{equation}

 \begin{equation}
     b_{us} = \hspace{0.3cm} b_{ds} \hspace{0.3cm} = \left(\frac{m_u}{m_s}\right)b_{uu} = 68.27 MeV
 \end{equation}

\begin{equation}
     b_{uc} = \hspace{0.3cm} b_{dc} = \hspace{0.3cm} \left(\frac{m_u}{m_c}\right)b_{uu} = 21.52 MeV
 \end{equation}

\begin{equation}
     b_{ss} = \hspace{0.3cm} \left(\frac{m_u}{m_s}\right)^2 b_{uu} =  45.63 MeV
 \end{equation}

\begin{equation}
     b_{sc} = \hspace{0.3cm} \left(\frac{m_u^2}{m_s m_c}\right) b_{uu} = 14.45 MeV
\end{equation}

\begin{equation}
     b_{cc} = \hspace{0.3cm} \left(\frac{m_u}{ m_c}\right)^2 b_{uu} = 4.5 MeV
 \end{equation}
Using these values of hyperfine interaction terms $b_{ij}$ and quark masses, effective quark masses for $J^P = 5/2^-$ are given as: \\
\\
i) For ($s$ = 0) decuplet pentaquarks,
\begin{equation}
    m_u^* = \hspace{0.3cm} m_d^* = 392.793 MeV
    \end{equation}
    \begin{equation}
    m_c^* = \hspace{0.3cm} m_{\Bar{c}}^* = 1718.63 MeV
    \end{equation}
\\
(ii) For ($s$ = 1) decuplet pentaquarks,
\begin{equation}
    m_u^* = \hspace{0.3cm} m_d^* = 388.62 MeV
    \end{equation}
    \begin{equation}
        m_s^* = \hspace{0.3cm} 559.68 MeV
    \end{equation}
    \begin{equation}
    m_c^* = \hspace{0.3cm} m_{\Bar{c}}^* = 1717.75 MeV
    \end{equation}
\\

(ii) For ($s$ = 2) decuplet pentaquarks,
\begin{equation}
    m_u^* = \hspace{0.3cm} m_d^* = 384.448 MeV
    \end{equation}
    \begin{equation}
        m_s^* = \hspace{0.3cm} 556.85 MeV
    \end{equation}
    \begin{equation}
    m_c^* = \hspace{0.3cm} m_{\Bar{c}}^* = 1716.87 MeV
    \end{equation}
\\

(iV) For ($s = 3$) decuplet baryons,
    \begin{equation}
        m_s^* = \hspace{0.3cm} 554.02 MeV
    \end{equation}
    \begin{equation}
    m_c^* = \hspace{0.3cm} m_{\Bar{c}}^* = 1715.98 MeV
    \end{equation}
\\
 There is no orbital excitation included for these states and only ground state magnetic moments have been calculated. The mass spectrum of hidden-charm decuplet of pentaquarks for $J^P = 5/2^-$ is listed in Table \ref{tab:5} and compared with the masses obtained by GR mass formula formalism and the multiquark color flux-tube model \cite{isomers}. \\
\subsection{Screened Charge Scheme}
As the quark masses get modified due to interaction between neighboring quarks, one can speculate that the quark's charge inside the pentaquark may also experience similar effects. There is a linear dependency of effective charge on the charge of shielding quarks. Effective charge of a quark in exotic baryon P$(a, b, c, d, e, f)$ is defined as:
\begin{equation}
    e_a^P = e_a + \alpha_{ab} e_b + \alpha_{ac} e_c + \alpha_{ad} e_d + \alpha_{af} e_f
\end{equation}
Where $e_a$ is the screened charge of $a$ quark due to neighboring quarks. By considering the isospin symmetry, we take
\begin{align}
 \alpha_{ab} = \alpha_{ba}, \hspace{0.3cm} \alpha_{us} =  \alpha_{ds}= \alpha, \hspace{0.3cm} \alpha_{ss} = \gamma  
\end{align}
In a similar manner, the charm sector
\begin{align}
    \alpha_{uc} = \alpha_{dc} = \beta, \hspace{0.3cm} \alpha_{sc} = \delta^{'}, \hspace{0.3cm} \alpha_{cc} = \gamma^{'}
\end{align}
and for the bottom sector,
\begin{align}
    \alpha_{ub} = \alpha_{db} = \beta^{''}, \hspace{0.3cm} \alpha_{sb} = \delta^{''}, \hspace{0.3cm} \alpha_{cb} = \gamma^{''}, \hspace{0.3cm} \alpha_{bb} = \xi
\end{align}
the parameter $\alpha_{ij}$ is defined as:
\begin{equation}
    \alpha_{ij} = \mid{\frac{m_i - m_j}{m_i + m_j}}\mid \times \delta
\end{equation}
here $\delta$  is equal to 0.81, by substituting effective charge values in the magnetic moment operator, we get:
\begin{equation}
\mu = \sum_i \frac{e_i^P}{2  m_i^{eff}}  \sigma_i
\label{MM}
\end{equation}
The expressions for magnetic moments of decuplet of hidden charm pentaquarks are written in Table \ref{tab: expressions} and the obtained numerical results from Eq.\eqref{MM} are written in Table \ref{tab: magnetic moments} and will be helpful for future experimental studies. In the next section, the production mechanism of pentaquark states has been studied via bottom baryon decay channels.
\begin{table}[]
    \centering
    \begin{tabular}{cc}
    \hline
    \hline
       Decuplet States  & \hspace{0.3cm} Effective mass Scheme  \\
       \hline
        $P_c^{++(0)}$  & \hspace{0.3cm} 3 $\mu_u^{eff}$ + $\mu_c^{eff}$ + $\mu_{\Bar{c}}^{eff}$
\\
        $P_c^{+(0)}$ & \hspace{0.3cm} 2$\mu_u^{eff}$ + $\mu_d^{eff}$ + $\mu_c^{eff}$ + $\mu_{\Bar{c}}^{eff}$ \\
        $P_c^{0(0)}$ & \hspace{0.3cm} $\mu_u^{eff}$ + 2$\mu_d^{eff}$ + $\mu_c^{eff}$ + $\mu_{\Bar{c}}^{eff}$ \\
        $P_c^{-(0)}$ & \hspace{0.3cm} 3$\mu_d^{eff}$ + $\mu_c^{eff}$ + $\mu_{\Bar{c}}^{eff}$  \\
        $P_c^{+(1)}$ & \hspace{0.3cm} 2$\mu_u^{eff}$ + $\mu_s^{eff}$ + $\mu_c^{eff}$ + $\mu_{\Bar{c}}^{eff}$  \\
        $P_c^{0(1)}$ & \hspace{0.3cm} $\mu_u^{eff}$ + $\mu_d^{eff}$ + $\mu_s^{eff}$ + $\mu_c^{eff}$ + $\mu_{\Bar{c}}^{eff}$ \\
        $P_c^{-(1)}$ & \hspace{0.3cm} 2$\mu_d^{eff}$ + $\mu_s^{eff}$ + $\mu_c^{eff}$ + $\mu_{\Bar{c}}^{eff}$   \\
        $P_c^{0(2)}$ & \hspace{0.3cm}  $\mu_u^{eff}$ + 2$\mu_s^{eff}$ + $\mu_c^{eff}$ + $\mu_{\Bar{c}}^{eff}$  \\
        $P_c^{-(2)}$ & \hspace{0.3cm} $\mu_d^{eff}$ + 2$\mu_s^{eff}$ + $\mu_c^{eff}$ + $\mu_{\Bar{c}}^{eff}$  \\
        $P_c^{-(3)}$ & \hspace{0.3cm}  3$\mu_s^{eff}$ + $\mu_c^{eff}$ + $\mu_{\Bar{c}}^{eff}$ \\
        \hline
        \hline
    \end{tabular}
    \caption{Expressions for the magnetic moments of ($J^P =
5/2^-$) pentaquarks using effective quark masses (in $\mu_N$ ).}
  \label{tab: expressions}
\end{table}

\begin{table*}
    \centering
    \begin{tabular}{ccccc}
    \hline
    \hline
     State & \hspace{0.3cm} Quark Content & \hspace{0.3cm} Effective mass scheme  & \hspace{0.3cm} Screened Charge Scheme & \hspace{0.3cm} Effective mass + Screened Charge Scheme \\
     \hline
$P_c^{++(0)}$ & \hspace{0.3cm} $uuuc\Bar{c}$ & \hspace{0.3cm} 4.78 & \hspace{0.3cm} 6.34 & \hspace{0.3cm} 5.93\\
      
$P_c^{+(0)}$ & \hspace{0.3cm} $uudc\Bar{c}$ & \hspace{0.3cm} 2.39 & \hspace{0.3cm} 3.17 & \hspace{0.3cm} 2.97 \\
       
$P_c^{0(0)}$ & \hspace{0.3cm} $uddc\Bar{c}$ & \hspace{0.3cm} 0 & \hspace{0.3cm} 0 & \hspace{0.3cm} 0 \\
         
 $P_c^{-(0)}$ & \hspace{0.3cm} $dddc\Bar{c}$ & \hspace{0.3cm} -2.39 & \hspace{0.3cm} -3.17 & \hspace{0.3cm} -2.96\\
         
 $P_c^{+(1)}$ & \hspace{0.3cm} $uusc\Bar{c}$ & \hspace{0.3cm} 2.66 & \hspace{0.3cm} 3.59 & \hspace{0.3cm} 3.37 \\
          
 $P_c^{0(1)}$ & \hspace{0.3cm} $udsc\Bar{c}$ & \hspace{0.3cm} 0.246 & \hspace{0.3cm} 0.14 & \hspace{0.3cm} 0.12\\
           
 $P_c^{-(1)}$ & \hspace{0.3cm} $ddsc\Bar{c}$& \hspace{0.3cm} -2.17 & \hspace{0.3cm} -3.30 & \hspace{0.3cm} -3.14 \\
            
$P_c^{0(2)}$ & \hspace{0.3cm} $ussc\Bar{c}$ & \hspace{0.3cm} 0.50 & \hspace{0.3cm} 0.74 & \hspace{0.3cm} 0.68 \\
             
$P_c^{-(2)}$ & \hspace{0.3cm} $dssc\Bar{c}$& \hspace{0.3cm} -1.94 & \hspace{0.3cm} -2.98  & \hspace{0.3cm} -2.87 \\
              
$P_c^{-(3)}$ & \hspace{0.3cm} $sssc\Bar{c}$& \hspace{0.3cm} -1.69 & \hspace{0.3cm} -2.20 & \hspace{0.3cm} -2.15\\
               \hline
               \hline
    \end{tabular}
    \caption{Magnetic moments of hidden-charm decuplet of pentaquarks having $J^P = 5/2^-$ using the effective mass scheme, screened charge scheme, and effective mass and screened charge schemes together.}
    \label{tab: magnetic moments}
\end{table*}
   
\section{Numerical Analysis}
\subsection{Bottom Baryon Decay Channels With Intermediate Pentaquark States}
\begin{figure}[h]
    \centering
    \includegraphics[width=0.8\linewidth]{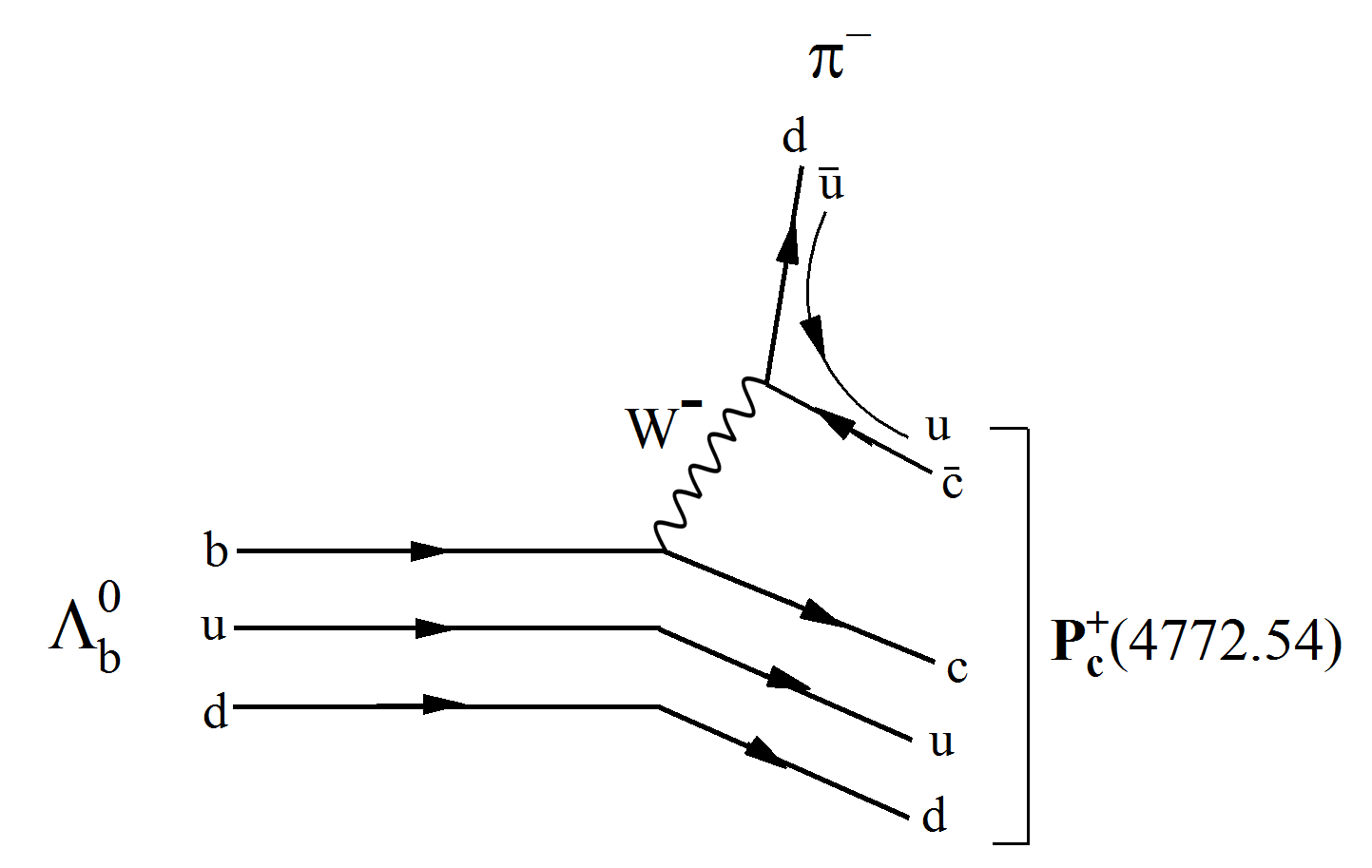}
     \caption{ $\Lambda_b^0$ decay to $P_c^{+(0)}$ and $\pi^-$ states,  $P_c^{+(0)}$ is a charged pentaquark state, and the member of isospin quartet with hypercharge $Y$ = 1. }
    \label{fig:2}
\end{figure}

As LHCb produces a large number of bottom baryons in $pp$ collisions, therefore, we predicted the possible bottom baryon decay channels for the production of hidden-charm pentaquark states. By considering these pentaquark assignments as genuine states, decay channels are described below in detail. \\
 Starting with the $P_c^{++(0)}$ state, which is part of an isospin quartet. One possible decay channel to observe this state could be:
\begin{equation}
\Sigma_b^+ \rightarrow P_c^{++(0)} + \pi^- , \hspace{0.5cm} P_c^{++(0)} \rightarrow  \Delta^{++} + J/\psi
\end{equation}
Production of $P_c^{++(0)}$ state from bottom baryon decay of $\Sigma_b^+$ involves the creation of $u\overline{u}$ in a vacuum. Now, one of its isospin partner, $P_c^{+(0)}$ can be observed in the following process:
\begin{equation}
\Lambda_b^0 \rightarrow P_c^{+(0)} + \pi^- , \hspace{0.5cm} P_c^{+(0)} \rightarrow  \Delta^{+} + J/\psi
\end{equation}
The corresponding Feynman diagram for this decay is shown in Figure \ref{fig:2}.
The other member of the isospin quartet, $P_c^{0(0)}$, can also be observed in the bottom baryon decay channel of $\Lambda_b^0$ as:
\begin{equation}
\Lambda_b^0 \rightarrow P_c^{0(0)} + \pi^0 , \hspace{0.5cm} P_c^{0(0)} \rightarrow  \Delta^0 + J/\psi
\end{equation}
The difference between the two suggested decay modes for $\Lambda_b^0$ lies in the final state. In the decay of $P_c^{+(0)}$ state, a $u\bar{u}$ is created from the vacuum, whereas, for $P_c^{0(0)}$ decay, a $u\overline{u}$ pair is replaced by the $d\overline{d}$ pair. Now, in the case of the production of $P_c^{-(0)}$ state, it can be seen in one of the decay channels of $\Xi_b^0$:
\begin{equation}
\Xi_b^0 \rightarrow P_c^{-(0)} + \overline{K}^0 , \hspace{0.5cm} P_c^{-(0)} \rightarrow  \Delta^- + J/\psi
\end{equation}
here, in the production of $P_c^{-(0)}$ state, a $d\overline{d}$ pair is created into the vacuum.\\
For other members of the decuplet i.e those with strangeness ($P_c^{+(1)}$, $P_c^{0(1)}$, $P_c^{-(1)}$) forms an isospin triplet. Starting with the production of $P_c^{+(1)}$, let us consider this decay of $\Lambda_b^0$ baryon:
\begin{equation}
\Lambda_b^0 \rightarrow P_c^{+(1)} + \pi^- , \hspace{0.5cm} P_c^{+(1)} \rightarrow  \Sigma^{*+} + J/\psi
\label{deltaplus}
\end{equation}
Further, the production mechanism of $P_c^{0(1)}$ can be observed in the bottom baryon decay channel of $\Lambda$ baryon, which further decays to $\Sigma^{*0}$ and $J/\psi$ state as:
\begin{equation}
\Lambda_b^0 \rightarrow P_c^{0(1)} + K^0 , \hspace{0.5cm} P_c^{0(1)} \rightarrow  \Sigma^{*0} + J/\psi
\label{deltazero}
\end{equation}
The two decay processes mentioned in eq. \eqref{deltaplus} and \eqref{deltazero} are different because, in the production of $P_c^{+(1)}$, the $u\overline{u}$ pair is created in a vacuum, and in the case of $P_c^{0(1)}$ production, a $s\overline{s}$ pair is created. Now, the third particle of the isospin triplet, $P_c^{-(1)}$ state, can be produced in the following decay process:
\begin{equation}
\Xi_b^0 \rightarrow P_c^{-(1)} + \pi^0 , \hspace{0.5cm} P_c^{-(1)} \rightarrow  \Sigma^{*-} + J/\psi
\label{deltaminus}
\end{equation}
This production of $P_c^{-(1)}$ from the bottom baryon decay of $\Xi_b^0$ consists of the $d\overline{d}$ pair in a vacuum. Now, we discuss the doubly strange pentaquarks forming an isospin doublet ($P_c^{0(2)}$, $P_c^{-(2)}$). Firstly, the production of $P_c^{0(2)}$ state by one of the possible decay channel of $\Omega_b^0$ baryon as:
\begin{equation}
\Omega_b^0 \rightarrow P_c^{0(2)} + \pi^- , \hspace{0.5cm} P_c^{0(2)} \rightarrow  \Xi^{*0} + J/\psi
\label{caszero}
\end{equation}
Also, for the case of it's isospin partner $P_c^{-(2)}$, it can be seen via following decay process:
\begin{equation}
\Xi_b^0 \rightarrow P_c^{-(2)} + K^0 , \hspace{0.5cm} P_c^{-(2)} \rightarrow  \Xi^{*-} + J/\psi
\label{casminus}
\end{equation}
Decay channels are written in eq. \eqref{caszero} and \eqref{casminus} are different in the quark-antiquark pair production in a vacuum. In $P_c^{0(2)}$ production, a $u\overline{u}$ pair is produced in vacuum while in  case of $P_c^{-(2)}$, a $s\overline{s}$ pair is created in the vacuum. At last, we talk about the production mechanism of triply strange singlet pentaquark states, $P_c^{-(3)}$, which is produced in the decay process:
\begin{equation}
\Omega_b^- \rightarrow P_c^{-(3)} + K^0 , \hspace{0.5cm} P_c^{-(3)} \rightarrow  \Omega^{-} + J/\psi
\end{equation}
This $\Omega_b^-$ decay to $P_c^{-(3)}$ state contains the production of $s\overline{s}$ pair in vacuum.

\section{Conclusion}
The vast upcoming data of hidden-charm pentaquarks at LHCb Collaborations served as the inspiration for this work. By using the multiquark approach, a spectroscopic analysis of the hidden-charm pentaquarks is carried out by classifying them in the decuplet of SU(3) representation with spin-parity $J^P = 5/2^-$. The allowed values of quantum numbers for the hidden-charm decuplet like isospin, and hypercharge are $I = 0, \frac{1}{2}, \frac{3}{2}, 1$ and $Y = 0, \pm 1, -2$ respectively. Using these allowed values of quantum numbers, we calculated the masses for each member of the decuplet using an extension of the Gursey-Radicati mass formula and the effective mass scheme and a comparison of the mass spectra is carried out with the available theoretical data. Also, by using the effective mass and screened charge scheme, we calculated the magnetic moment of the hidden-charm decuplet of pentaquarks which will help us to explore the inner structure of pentaquarks. Effective mass for each quark has been calculated using the one gluon exchange scheme. Also, by using the screened charge scheme, we calculated the screened charge of hidden-charm pentaquarks to calculate the magnetic moments. Table \ref{tab: magnetic moments} consists of the magnetic moments of hidden-charm decuplet of pentaquarks using the effective mass, screened charge, and effective mass and screened charge together. Further, the LHCb observes a large number of bottom baryon decays which may produce intermediate pentaquark states as seen in $\Lambda_b^0 \rightarrow J/\psi pK^-$ channels. We suggested the possible bottom baryon decay channels producing intermediate pentaquark states studied in this work. Thus, our analysis may provide helpful information for observing the suggested decay channels for hidden-charm states in future experimental studies. 

\section*{Acknowledgment}
The authors thankfully acknowledge the financial support from the Department of Science and Technology, New Delhi. 
Project No. CRG/2019/005159. File No. (SERB/F/9119/2020).

\bibliography{sample}% Produces the bibliography via BibTeX.

\end{document}